# Energy deposition by heavy ions: Additivity of kinetic and potential energy contributions in hillock formation on $CaF_2$

Y. Y. Wang[1], C. Grygiel[2], C. Dufour[2], J. R. Sun[1], Z. G. Wang[1], Y. T. Zhao[1], G. Q. Xiao[1], R. Cheng[1], X. M. Zhou[1], J. R. Ren[1], S. D. Liu[1], Y. Lei[1], Y. B. Sun[1], R. Ritter[3], E. Gruber[3], A. Cassimi[2], I. Monnet[2], S. Bouffard[2], F. Aumayr[3] and M. Toulemonde[2,*]

[1]*Institute of Modern Physics, Chinese Academy of Sciences, Lanzhou 730000, China*
[2]*CIMAP-GANIL, CEA-CNRS-ENSICAEN-Univ.of CAEN, F-14070 Caen Cedex 5, France*
[3]*Institute of Applied Physics, TU Wien, 1040 Vienna, Austria*
**Corresponding author:toulemonde@ganil.fr*

Abstract:

The formation of nano-hillocks on $CaF_2$ crystal surfaces by individual ion impact has been studied using medium energy (3 and 5 MeV) highly charged ions ($Xe^{19+}$ to $Xe^{30+}$) as well as swift (kinetic energies between 12 and 58 MeV) heavy ions. For very slow highly charged ions the appearance of hillocks is known to be linked to a threshold in potential energy while for swift heavy ions a minimum electronic energy loss is necessary. With our results we bridge the gap between these two extreme cases and demonstrate, that with increasing energy deposition via electronic energy loss the potential energy threshold for hillock production can be substantially lowered. Surprisingly, both mechanisms of energy deposition in the target surface seem to contribute in an additive way, as demonstrated when plotting the results in a phase diagram. We show that the inelastic thermal spike model, originally developed to describe such material modifications for swift heavy ions, can be extended to case where kinetic and potential energies are deposited into the surface.

The interaction of individual slow highly charged ions (HCI) in the keV energy regime with solid surfaces is able to induce surface modifications on a nanometric scale [1], which appear as either hillocks, pits or craters [2-7]. These modifications result from the deposition of potential energy carried by the HCI (i.e. the sum of binding energies of all missing electrons) into the target electronic system absorbing tens of keV/nm$^3$ within a few femtoseconds (fs) [8]. Similarly, hillocks created by swift heavy ions (SHI) with impact energies of some hundreds of MeV, appear at the surface for an energy density of typically 10 keV/nm$^3$ deposited by ion-electron collisions also in the fs time range [9-11]. This similarity between HCI and SHI was pointed out by Aumayr et al. [1] in order to establish a qualitative link between surface modifications by potential energy ($E_p$) or by electronic energy loss ($S_e$). Interpreting their results for $CaF_2$, El-Said et al. [2] were the first to suggest that hillock formation by slow HCI may be described by the inelastic thermal spike model [12] originally developed to predict material modifications induced by SHI. Along the lines of this model, the sharp threshold of potential energy for hillock formation by slow HCI was linked to a solid-liquid phase transition (nano-melting) [1,2]. In this study we establish the so far missing quantitative link between HCI and SHI induced nano-hillocks, by demonstrating that potential energy deposition and electronic energy loss act together in an additive way to induce surface modifications.

$CaF_2$ (111) crystal (Korth Crystal Company) were freshly cleaved in air before placing them in the high vacuum irradiation chamber. After irradiation, samples were imaged under ambient conditions by Atomic Force Microscopy (AFM) in tapping mode using either a NanoscopeIII (DI) in Lanzhou or a Cypher AFM (Asylum Research) in Vienna. Image analysis was done with the Gwyddion software [13]. Using statistical methods this code allows to determine the surface roughness and the number of hillocks per unit area for each sample. Freshly cleaved surfaces of $CaF_2$ are flat with a mean roughness of 0.07 nm and no changes are observed when inspected after 5 and 16 days. Also the irradiated surface keeps the same roughness in the time within the experimental errors.

Medium energy irradiations using Xe ions were carried out on the 320 kV ECR platform for highly charged ions physics research at IMP (Lanzhou). The fluence was about $5\times10^{10}$ $Xe^{q+}/cm^2$ per sample. $CaF_2$ surfaces have been irradiated at two kinetic energies (3 and 5 MeV) by $Xe^{q+}$ with charge state q between 19 and 26. All projectile charge states used are therefore below the charge state threshold for hillock formation reported to be $Xe^{28+}$ for very low kinetic energy (0.004 MeV) [2]. In addition irradiations with $Xe^{30+}$ ions have been performed at three kinetic energies (0.54, 3 and 5 MeV).

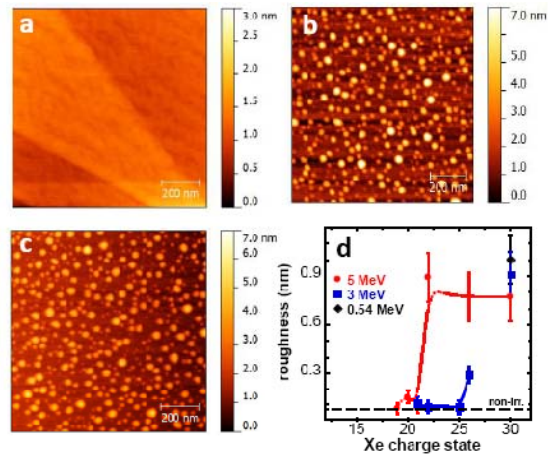

FIG. 1 (color online). $CaF_2$ surfaces irradiated by 5 MeV Xe ions with a fluence of 500 ions/$\mu m^2$. All the images show an area of $1\times1$ $\mu m^2$. (a) presents a surface irradiated by $Xe^{21+}$ with no hillocks visible; (b) one by $Xe^{22+}$ with ~520 hillocks; and (c) one by $Xe^{30+}$ with ~410 hillocks. (d) presents the evolution of the surface roughness versus charge state for non-irradiated, 5, 3 and 0.54 MeV Xe ion irradiations.

Fig. 1 shows surfaces of samples irradiated by $Xe^{21+}$, $Xe^{22+}$, and $Xe^{30+}$ ions with a kinetic energy of 5 MeV. For $Xe^{21+}$ (Fig. 1a) no hillocks are observable and the irradiated sample stays flat with a mean roughness of 0.07 nm. The same is true for surfaces irradiated by ions in lower charge states ($Xe^{20+}$, $Xe^{19+}$; not shown). However, nano-hillocks clearly appear on surfaces after irradiation by $Xe^{22+}$ (Fig. 1b), $Xe^{26+}$ (not shown) and $Xe^{30+}$ (Fig. 1c).

For these cases the number of hillocks per area corresponds to the fluence of incident ions within our experimental errors. For irradiations performed with a kinetic energy of 3 MeV, nanoscale hillocks appear on the $CaF_2$ surface only for charge states equal and larger than 26+. Samples irradiated by 3 MeV $Xe^{25+}$ ions (and ions in smaller charge states) do not show any hillocks, while for $Xe^{26+}$ and $Xe^{30+}$ hillocks are clearly visible.

To illustrate the surprisingly sharp but impact energy dependent transition from flat to nanostructured $CaF_2$ surfaces, the roughness of the irradiated samples as defined by the Gwyddion code [13], is plotted in Fig. 1d versus Xe charge states for different kinetic energies (i.e. 3 and 5 MeV; also showing the result of one irradiation performed at only 0.54 MeV). Right at the threshold (22+ for 5 MeV and 26+ for 3 MeV) the surface roughness increases significantly to values 4 – 10 times that of the non-irradiated sample.

In comparison, at a kinetic energy of 5 MeV the potential threshold for hillock formation is between $Xe^{21+}$ and $Xe^{22+}$, leading to a value of $E_p \approx 5.5$ keV [14], while at a kinetic energy of 3 MeV the potential threshold is situated between $Xe^{25+}$ and $Xe^{26+}$, leading to a value of $E_p \approx 8.5$ keV. Both values for the potential energy threshold measured in the MeV energy regime are considerably smaller than the one reported by El-Said et al. [2] in the keV energy regime ( between $Xe^{27+}$ and $Xe^{28+}$; equal to $E_p \approx 11.2$ keV).

With increasing kinetic energy, nuclear and electronic energy losses of the projectile ion evolve differently. For energies of 0.004 and 5 MeV, the nuclear energy loss is nearly constant varying only from 1.15 keV/nm to 1 keV/nm respectively [15]. Such a 15% decrease of the nuclear energy loss within increasing kinetic energy cannot account for the observed variation in potential energy threshold by 50% from 11.2 to 5.5 keV. But in the same energy range, the electronic energy loss increases from 0.02 to 1.48 keV/nm. Since by electronic stopping $S_e$ the electronic system of the target is excited in a similar way as by deposition of potential energy of a HCI [1], we conclude that a decrease of the potential energy deposition (decreasing $E_p$ threshold) can be counter-balanced by an increased energy deposition via $S_e$.

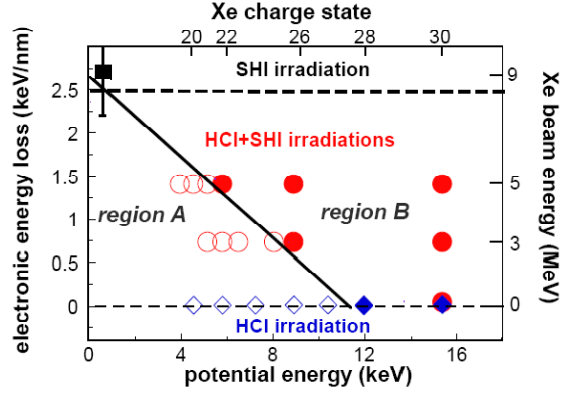

FIG. 2 (color online). *Hillock formation on CaF$_2$ as a function of electronic energy loss deposited (x-axis) and potential energy carried (y-axis) by the Xe ions. The corresponding kinetic energies and Xe charge states are given on the opposite axis, respectively. Open symbols represent cases where no hillocks have been found (region A), while full symbols mean appearance of hillocks (region B). The blue diamond points at energies of ~0.003 and ~0.3 MeV are from [1, 2]. Red circle points are from the present experiments. The black square point at 9.2 MeV just above the dotted line is our measured threshold energy of hillock appearance by SHI (see below).*

In Fig. 2 we have summarized the results of our investigations in a "phase diagram" with $S_e$ and $E_p$ as state variables. Cases, where hillocks have been found after irradiation are displayed by full symbols (region B), while open symbols represent irradiations where the CaF$_2$ surface stayed flat and showed no effect of the irradiation (region A). Also data from slow HCI impact taken from refs [1, 2] and our results from swift heavy ion impact (see below) have been included in this plot. The $S_e$ values (0.72 keV/nm at an energy of 3 MeV and 1.48 keV/nm for 5 MeV) were calculated with the CasP code [16] since it predicts more realistic electronic energy loss values [17] as compared to SRIM [15] in this low MeV energy regime. The border line between region A (no hillocks) and region B (hillocks) in the $S_e - E_p$ diagram (Fig. 2) follows to a good approximation a straight line with negative slope, thus pointing to an additive contribution of $S_e$ and $E_p$ to nano-hillock formation on the surface of $CaF_2$. A decrease of $E_p$, deposited in the target clearly can be compensated by an increase of $S_e$ and vice versa. Without any additional potential energy a minimum electronic energy loss of $S_e = 2.65 \pm 0.40$ keV/nm is necessary to create hillocks, as can be seen from the linear extrapolation of the border line in Fig. 2 to $E_p=0$. If, on the other hand, the electronic energy loss of the projectile is negligibly small, the threshold in $E_p$ is close to 12 keV.

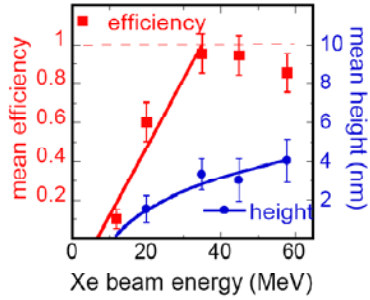

FIG. 3 (color online).  Mean height (blue circles) and efficiency (red squares) for hillock formation by swift heavy ions versus Xe ion beam energy.

We note, that the $S_e$ threshold for SHI (i.e. for $E_p=0$) of 2.65 keV/nm in fig. 2 is lower by a factor of 2 than the $S_e$ threshold (~5 keV/nm) reported previously by SHI in the GeV regime [18]. Such a decrease can in principle result from the so-called "velocity effect": For SHI impact on a surface with different kinetic energy but same electronic energy loss, the energy deposited into the electronic system at lower ion velocity is more efficiently transferred to the lattice atoms than at higher energies [19]. However, in the past it has been argued that $CaF_2$ should not be sensitive to the velocity effect [20]. To settle this dispute we have re-measured the kinetic energy threshold for hillock creation by SHI in the MeV energy regime using Xe ions from the IRRSUD beam line of the GANIL facility [21]. The initial beam energy was 92 MeV and using Al foil degraders, Xe energies ranging between 12 and 58 MeV could also be realized. The fluences applied were $5 \times 10^9$ Xe/cm$^2$ and $1.5 \times 10^{10}$ Xe/cm$^2$, respectively. Using AFM in tapping mode [11], the surface of each sample was analyzed to extract the number of hillocks per cm$^2$ and the mean hillock height. The mean hillock height as well as the mean efficiency for hillock production (number density of hillocks divided by incident ion fluence) were deduced for the irradiations at the two fluences. In Fig. 3 the measured efficiencies and heights are plotted versus Xe beam energy. For extrapolation to zero efficiency a linear fit to the points at energies of 12, 20 and 35 MeV was applied, leading to threshold beam energy of 6.7 MeV. Given that the number of hillocks at a beam energy of 12 MeV Xe was low due to the small efficiency, the mean hillock height has large error bars and is not reported here. Since in this kinetic energy regime the $S_e$ varies as the square root of the beam energy [22] and the height should be proportional to $S_e$ [7], the height values were fitted with a square root law. Extrapolation of such a fit to zero height results in a beam energy threshold of 11.7 MeV (Fig. 3). Combining these two values, leads to a threshold energy of 9.2 ± 2.5 MeV for hillock formation in our MeV energy region. Using CasP code [16] this translates to an electronic energy loss threshold of $S_e$ = 2.7 ± 0.5 keV/nm in close agreement in agreement with the extrapolation of the border line to $E_p = 0$ in Fig. 2. Our results also confirm that $CaF_2$ is indeed sensitive to velocity effect [23] and $S_e$ threshold values in the (low velocity) MeV region differ from threshold values in the high velocity GeV region.

The additivity between $S_e$ and $E_p$ derived from Fig. 2 can be expressed in a quantitative formula taking into account that only a fraction F [24] of the $E_p$ is deposited in a depth d near the surface [25].

$$S_e + \frac{F}{d} E_p = const. = 2.7 \ keV/nm \qquad (1)$$

F can be either derived from calorimetric measurements [26] or by measuring the energy carried away by emitted electrons [24]. From our results (equ. 1) a ratio of the depth d in nanometer to the fraction F can be derived:

$$\frac{d}{F} = 4.3 \pm 0.5 \text{ nm} \tag{2}$$

In the following we present an extension of the inelastic thermal spike (i-TS) model [27] originally developed to describe material modifications induced by electronic energy loss of SHI, to medium and low energy HCI where also potential energy is deposited. As a central assumption, the threshold for hillock formation is linked to a solid-liquid phase transition (nano-melting) [1]. The original i-TS model describes a transient thermal process based on heat transport equations that govern the heat diffusion in time and space (radial distance from the ion path) in the electronic and atomic subsystems and their coupling by the electron-phonon constant $g$. For SHI the two equations are solved numerically in cylindrical geometry. The initial energy distribution to the electrons is derived from Monte Carlo calculations and implemented as an analytical formula [28]. The electron-phonon coupling constant $g$ is linked to the electron-phonon mean free path $\lambda$ [27] that defines the mean length of energy diffusion on the electrons before its transfer to the atoms. The model was successfully applied for SHI in the GeV energy regime on $CaF_2$ [18] with $\lambda$ equal to 3.8 nm, assuming that tracks and hillocks appear, if the energy $E_m$ = 0.58 eV/atom to melt the material is surpassed. Such calculated electronic energy loss threshold for material modification is predicted to be 5 keV/nm in close agreement with experiments in the GeV energy regime [18]. Applying the i-TS model to heavy ions in the MeV regime [19] a lower threshold of 2.6 ± 0.5 keV/nm for hillock formation is calculated, which is in perfect agreement with our experimental results presented in Fig. 2. The "theoretical" error of 0.5 keV/nm basically reflects the uncertainties in the initial energy distribution of the electrons.

In order to take into account that the energy given to the electrons by the potential energy and the electronic energy loss is not constant along the ion path, the 3D i-TS model developed for metallic materials [29] has been adapted for insulators following the assumptions made for the electron subsystem as already described [27]. It is also assumed that a fraction F of the potential energy is homogeneously deposited into a cylindrical volume below the surface [27] which is characterized by a depth d and a radius $R_p$. For the numerical calculation, it is of course necessary to estimate F, d and $R_p$. The radial distribution of potential energy has been estimated by Lemell et al. [30], showing that the most efficient electrons are the ones with energy lower than 400 eV. These electrons are confined in a cylinder radius $R_p$ between 0.5 and 2 nm. In the following, we therefore use $R_p$ = 1 nm to make quantitative calculations. Up to the depth d below the surface $S_e$ and the fraction $F*E_p$ are additively deposited, while for larger depth only $S_e$ is the only source.

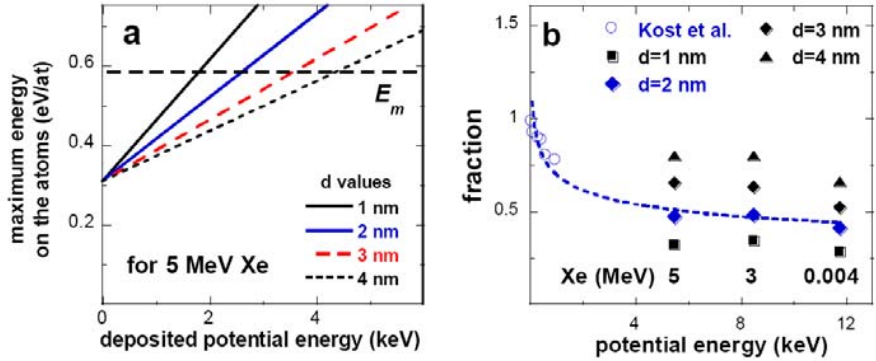

FIG. 4 (color online). (a) Maximum deposited energy per atoms for $Xe^{22+}$ projectile ions at 5 MeV impact energy as a function of the deposited potential energy. i-TS calculations for an initial radial energy distribution of 1 nm and four different values for the depth up to which potential energy is distributed (d=1, 2, 3 and 4 nm) are presented. A value of Em = 0.58 eV/at has to be surpassed to enable melting of the material. (b) Fraction F of deposited potential energy to reach the melting energy versus potential energy Ep. Open circles correspond to the measurements of Kost et al. [24]. The dotted line is a fit to our results for d =2 nm and the values of Kost et al. using a power law (c.f. text).

Fig. 4a compares the resulting energy $E_a$ transferred to an atom for 5 MeV Xe impact on $CaF_2$ to the necessary threshold energy $E_m$ for melting $CaF_2$, thus deriving the amount of $E_p$ deposited into a depth d necessary for melting, i.e. the fraction F of $E_p$ where $E_a$ equals $E_m$. For the different values of d (1, 2, 3 and 4 nm) corresponding ratios d/F are derived (i.e. 3.2, 4.4, 5.1 and 5.4 nm, respectively). Knowing from our experiment that F and d should be linked by equ. (2) with a value of 4.3 nm, leads to a plausible depth of d = 2 nm. These calculations are repeated for Xe ions at 3 MeV (our experiment) and for Xe at 0.004 MeV (El-Said et al. experiments [2]). The such derived fractions F of the potential energy deposited into the target is plotted in Fig. 4b versus the experimentally measured potential threshold energies with d as parameter varying between 1 and 4 nm. For d = 2 nm the deposited fraction of $E_p$ is about 45% nearly independent of the ion`s kinetic energy. In Fig. 4b our F values (for 2 nm) are compared to the measurements of Kost et al. [24] who have actually determined the complimentary fraction of potential energy (1-F) which is carried away by electron emission. Fitting the Kost et al. data points and the present F values by a power law, we find the fraction F decreases with increasing $E_p$. This surprising dependence of F on $E_p$ might, however, only reflect the fact that with increasing potential energy more energetic electrons are produced during the decay of the hollow atom which carries away an increasingly large fraction of the potential energy.

This work has been supported by the Major State Basic Research Development Program of China ('973' Program, Grant no. 2010CB832902), the National Natural Science Foundation of China (Grant Nos. 11275238, 11275005, 11105192, 91126011 and 11275241) and the Austrian Science Fund FWF (I 1114-N20). We would like to thank the staff from the 320 kV ECR platform at IMP-CAS, the IRRSUD line of the GANIL accelerator and CIMAP laboratory for their help during the experiments.